\documentclass[12pt,prl]{revtex4}
\usepackage{graphicx}
\usepackage{mathrsfs}
\usepackage{amsfonts}
\usepackage{amssymb}
\usepackage{bm}
\usepackage{mathtools}

\DeclareMathAlphabet{\mathpzc}{OT1}{pzc}{m}{it}

\begin{document}

\title{
\vspace{-2cm}\hfill{\normalsize NSF-KITP-11-173}\\[1cm]
From constrained stochastic processes to the nonlinear sigma
  model. Two old problems revisited}

\author{Franco Ferrari}
\affiliation{
 Institute of Physics and CASA*, University of Szczecin, Wielkopolska 15,
70451 Szczecin, Poland\\ and\\
Kavli Institute for Theoretical Physics, University of
  California, Santa Barbara, California 93106, USA
}

\begin{abstract}
In this work a method is presented to derive the generating functional
in path integral form for a system with an arbitrary number of
degrees of freedom and constrained by general conditions.
The method is applied to the case of the dynamics of an inextensible
chain subjected 
to external forces. Next, 
the generating functional of the inextensible chain is computed 
assuming that the interactions are switched off. 
Finally, the generating functional of a two dimensional nonlinear
sigma model with 
O(3) symmetry is derived exploiting its similarities with the model
describing 
the dynamics of the inextensible chain.
\end{abstract}

\maketitle

\section{Introduction}
Stochastic systems in the presence of constraints are studied in
connection with several physical problems \cite{marko1,nelson,
  marko2,liverpool,febbo,pieranski,wardetal,Mochizuki, 
  okano}. 
 In the absence of constraints, a very useful method in order to solve
stochastic differential equations like those of Fokker--Planck
consists in expressing their solution in path integral form
\cite{ZinnJustin,hochberg,damgaard,msr}. Once this 
has been done,  a rich choice of path integral techniques is available
for concrete calculations. The derivation of the path
integral 
solutions of the Fokker-Planck equations becomes
however complicated when constraints are added. 
 Various approaches have been proposed in
order to cope with constrained stochastic differential equations
\cite{Mochizuki, 
  okano, doiedwards,fixman, granek,
  blundell,Gomes,Muthukumar, Hinch, curtiss,liverpool-maggs,
  Doyle,article2, abraham,  montesi, klaveness, 
  butler, echenique, peters}, but the available path integral
formulations for statistical 
systems out of equilibrium
still remain awkward to be used in practical applications.

In this work a
method is presented to derive
for a system out of equilibrium and with an arbitrary
number of degrees of freedom
a path integral
 expression of the
generating functional of the relevant correlation
functions.
The starting point are the Langevin
equations describing the unconstrained system with the addition
general constraint conditions. Roughly speaking, the strategy
behind the method is to
introduce fictitious degrees of freedom and white noises in order to
allow the variables that should be constrained to fluctuate.
It turns out that the dynamics of this extended system can be described by
an enlarged set 
of overdamped 
Langevin equations, so that it is possible to construct the path integral
of the related generating functional by using standard techniques. The
original system of Langevin 
equations and of constraints is recovered in the limit in which
the friction coefficients of the fictitious degrees of freedom and
the widths of the fictitious noise distributions
approach zero. In this limit, in the path integral of the generating
functional appear functional Dirac delta functions whose role is to
impose the 
constraints. 
The proposed method is tested in
the case of a discrete 
inextensible chain, 
for which the generating functional of the relevant correlation
functions has already been derived in path integral form in
\cite{FePaVi} and has undergone several consistency checks \cite{FePa,FePy}.
A considerable advance with respect to \cite{FePaVi} is the inclusion
in the treatment of the external forces, that were previously  missing.
While with the usual approaches the
path integral of the generating functional
contains extra degrees of freedom  (the Lagrange multipliers)
\cite{Mochizuki,okano,Klauder, gozzi} or,
alternatively, is complicated by functional
determinants of block matrices
\cite{Mochizuki,okano,itzyksonzuber}, in the present approach
the functional 
determinants are relatively simple and can be computed analytically.

Even with the simplest possible formulation, the treatment of
system with constraints is complicated by
the fact that
constraints involve nonlinear functions which
do not contain small or large parameters.
In this situation, any approximation
implies a change of the
imposed conditions. 
For instance, within the present approach the main problem are
the functional delta functions imposing the
constraints.
They make the computation of the generating functional extremely
complicated. 
Usually,
the constrained path integral is reduced to a gaussian one at the
 price of drastic simplifications \cite{FePaVi,EdwGoo,EdwGoo2,EdwGoo3}.
In the second part of this article it is shown in the case of the
inextensible chain 
that it is possible to  go beyond the
gaussian approximation. To this
purpose, the 
path integral of the generating functional of an inextensible
chain is rewritten as a product of two path 
integrals describing two systems interacting with an auxiliary
field.
One of the two path
integral
 is gaussian, while the
second, which contains the Dirac delta function coming from the
constraints,
 may be 
computed  by performing a series expansion
in
powers of the auxiliary field. The series converges provided a
parameter which is the analog of the Planck 
constant in quantum mechanics is small. 
An exact expression of
the  terms of any order appearing in the  series is derived. 
The generating functional of these terms is computed in closed form.

The path integral of a  inextensible
chain in three dimensions is very similar to that of an $O(3)$
nonlinear sigma model \cite{gellmann} in two dimensions. For this
reason, it is 
possible to apply also to this type of  sigma models
 the same 
method used to evaluate the generating functional of the inextensible chain. 

The material presented in this paper is divided as follows.
In the next Section, a method for constructing the generating
functional of a general constrained stochastic system is
presented. As a concrete example, we apply this method to the case of
a 
inextensible chain consisting of a set of beads connected together by
massless segments of fixed length. Next, we compute the generating
functional in the limit in which the chain becomes continuous.
In order to do that, it is necessary to perform a path integral that is
complicated by the presence of a functional Dirac delta function
related to the constraints. The path
integration is simplified by  the introduction of suitable auxiliary
fields. At the end, a
nonperturbative expression of the generating functional is obtained.
Finally, a similar method is exploited to compute the generating
functional of the correlation functions of a two dimensional nonlinear
sigma model with $O(3)$ group of symmetry.

\section{Constrained stochastic processes}
Let $\{\boldsymbol R\}=\boldsymbol R_1(t),\ldots,\boldsymbol R_N(t)$
denote the radius vectors of a set of $N$ particles in three
dimensions. The positions of the particles are constrained by the
conditions:
\begin{equation}
C_\alpha(\{\boldsymbol R\})=0\qquad\qquad
\alpha=1,\ldots,M<3N\label{constraints} 
\end{equation}
Their motion is subjected to the forces $\boldsymbol F_i=\boldsymbol
f+\boldsymbol \nu_i$ for $i=1,\ldots,N$. Apart from a rescaling by a
dimensional constant, namely the inverse of the friction coefficient, 
the $\boldsymbol
f_i=\boldsymbol f_i(\{\boldsymbol R\})$ are external forces, while the
$\boldsymbol \nu_i$'s represent random forces with gaussian
distribution, whose properties are determined by the following
correlation functions:
\begin{eqnarray}
\langle
\boldsymbol \nu_i(t)
\rangle&=&0\\
\langle
\nu_{i,a}(t)\nu_{j,b}(t')
\rangle&=&2D\delta_{ij}\delta_{ab}\delta(t-t')
\end{eqnarray}
Here $a,b=1,2,3$ label the spatial components and $D$ 
characterizes the width of the noise distribution.
Assuming that the particles are fluctuating in a viscous solution,
their motion will be described by the overdamped stochastic equation:
\begin{equation}
\dot{\boldsymbol R}_i=\boldsymbol f_i+\boldsymbol \nu_i\qquad\qquad
i=1,\ldots,N \label{Langevinequation}
\end{equation}
to which one has to add the conditions (\ref{constraints}).
Following a standard notation, in Eq.~(\ref{Langevinequation}) it has
been put $\dot{\boldsymbol R}_i(t)=\frac{\partial\boldsymbol
  R_i(t)}{\partial t}$.
Before passing from Eqs.~(\ref{constraints}) and (\ref{Langevinequation})
to a path integral formulation of the generating
functional of the correlation functions 
\begin{equation}
G_{i_1,\ldots,i_n,a_1,\ldots,a_n}(t_1,\ldots,t_n)=\langle
R_{i_1,a_1}(t_1),\ldots,R_{i_n,a_n}(t_n)
\rangle
\end{equation}
we briefly present a strategy explained in \cite{Chaichanbook} in
order to construct the generating functional in the simple case of the
Langevin equation of a particle moving along the
$x-$axis:
\begin{equation}
m\ddot{x}+\eta\dot{x}-f=\nu\label{equ1}
\end{equation}
with $\eta$ being the friction coefficient. 
Due to the inertial term, Eq.~(\ref{equ1})
is of second order in time.
The idea of
\cite{Chaichanbook} is to transform it in a system of
two equations of the first order by introducing the new variable 
\begin{equation}
v=\dot{x}\label{equ2}
\end{equation}
In the enlarged two dimensional space $(x,v)$, Eq.~(\ref{equ2}) looks like
a constraint which can be imposed
\'a la Parisi-Wu by adding a fictitious random force $\tilde \nu$ with
a gaussian noise distribution characterized by the standard deviation
$\tilde D$. In this way, one 
obtains from Eq.~(\ref{equ1})  two overdamped
Langevin equations:
\begin{eqnarray}
m\dot{v}+\eta v-f&=&\nu\label{equ1new}\\
\dot{x}-v&=&\tilde\nu\label{equ2new}
\end{eqnarray}
Starting from the above system of equations, the construction of
the generating functional of the correlation functions is
straightforward.
Unfortunately, this
method for fixing the constraints cannot be applied to our
case without changes because of two relevant differences. First, the constraint
(\ref{equ2}) is imposed on a new variable, the velocity $v$, that was
not present in the original Langevin equation (\ref{equ1}). For this
reason, there is since the beginning a clear distinction between the
physical and 
redundant degrees of freedom. In principle, it is possible to extract
also from Eqs.~(\ref{constraints}) and (\ref{Langevinequation}) a 
set of physical variables $u_1,\ldots,u_{3N-M}$ such that
$C_\alpha(
\boldsymbol R_1(u_1,\ldots,u_{3N-M}),\ldots, \boldsymbol R_N(u_1,\ldots,u_{3N-M})
)=0$ for $\alpha=1,\ldots,M$. The velocities
$v_1,\ldots,v_M$
 could then be identified
with the spurious degrees of freedom living in
the subspace of $\mathbb{R}^{3N}$ which is orthogonal to that
spanned by the $u$'s. The major difficulty of this program is that, in order
to 
determine the 
variables $u_1,\ldots,u_{3N-M}$, it is necessary to solve the relations
(\ref{constraints}). This task is in general not feasible. On the second
hand, Eq.~(\ref{equ2}) is already in the form of an overdamped
Langevin equation apart from the presence of the noise, while
Eqs.~(\ref{constraints}) are not. To eliminate these differences, we
introduce the new degrees of freedom $\xi_\alpha$ and new random forces
$\tilde\nu_\alpha$, $\alpha=1,\ldots,M$, characterized by the
correlation functions:
\begin{eqnarray}
\langle
\tilde{\nu}_\alpha(t)
\rangle&=&0\\
\langle
\tilde \nu_{\alpha}(t)\tilde\nu_{\beta}(t')
\rangle&=&2\tilde D_\alpha\delta_{\alpha\beta}\delta(t-t')
\end{eqnarray}
At this point, we replace Eqs.~(\ref{constraints}) and
(\ref{Langevinequation}) with the following set of Langevin equations:
\begin{eqnarray}
\dot{\boldsymbol R}_i&=&\boldsymbol f_i+\boldsymbol
\nu_i\label{Langevinequationsbis}\\ 
\eta_\alpha\dot{\xi}_\alpha+C_\alpha(\{\boldsymbol R\})&=&\tilde\nu_\alpha
\label{Langevinconstraints}
\end{eqnarray}
In the equilibrium limit $\tilde D_\alpha\to 0$ and by requiring that
the friction constants $\eta_\alpha$ vanish identically, the Langevin
equations (\ref{Langevinconstraints}) reduce to the original
constraints (\ref{constraints}).
The main advantage of having enlarged in this way the space of
variables is that now it is possible to apply
the standard procedure for constructing the expression of the generating
functional. First of all, we
introduce two sets of external currents $\boldsymbol J_i$,
$i=1,\ldots,N$ and $\tilde j_\alpha$, $\alpha=1,\ldots,M$. The
generating functional may be written as follows:
\begin{eqnarray}
Z[\boldsymbol J,\tilde j]&=&
\left[
\prod_{i=1}^N\prod_{\alpha=1}^M
\lim_{\tilde D_\alpha\to  0}
\lim_{\eta_\alpha\to 0}
\int{\cal D}\boldsymbol \nu_i{\cal  D}\tilde\nu_\alpha 
\right] 
\exp
\left\{-\int_0^{t_f}dt\left(
\sum_{i=1}^N\frac{\boldsymbol\nu_i^2}{4D}
+\sum_{\alpha=1}^M\frac{\tilde\nu_\alpha^2}{4\tilde  D_\alpha} 
\right)
\right\}
\nonumber\\
&\times&\exp\left\{
-\int_0^{t_f}dt\left[
\sum_{\alpha=1}^M\xi_\alpha\tilde j_\alpha+\imath\sum_{i=2}^N\boldsymbol
R_i\cdot(\boldsymbol J_{i} -\boldsymbol J_{i-1})
\right]\right\}\label{generatingfunctional}
\end{eqnarray}
In the above equation the quantity $t_f$ denotes the upper limit of
the time interval in which the evolution of the system is followed.
We stress the fact that the current term 
for the radius vectors 
$\imath\int_0^{t_f}dt\sum_{i=2}^N\boldsymbol
R_i\cdot(\boldsymbol J_{i} -\boldsymbol J_{i-1})$,
with $\imath$ denoting the imaginary unit, contains the discrete
derivative 
of the external currents $\boldsymbol J_i$ and not the current
itself. This definition of the
current term is a just matter of future convenience and does not
reduce the generality of our treatment.
Let us also note that in Eq.~(\ref{generatingfunctional}) the
$\xi_\alpha$'s and the $\boldsymbol R_i$'s represent respectively the
solutions of Eqs.~(\ref{Langevinconstraints}) and
(\ref{Langevinequationsbis}) and thus they depend on the noises
$\tilde\nu_\alpha$ and $\boldsymbol \nu_i$. Moreover, the $\boldsymbol
R_i(t)$'s satisfy the 
time boundary conditions:
\begin{equation}
\boldsymbol R_i(0)=\boldsymbol R_{0,i}\qquad\qquad\boldsymbol
R_i(t_f)=\boldsymbol R_{f,i} \label{boundaryconditions}
\end{equation}
Here the $\boldsymbol R_{0,i}$'s denote the fixed initial conformations of the
chain, while the $\boldsymbol R_{f,i}$'s describe the conformation at
the final 
time $t=t_f$.
Next, we perform in Eq.~(\ref{generatingfunctional}) the change of
variables
$\boldsymbol\nu_i, \tilde \nu_\alpha
\longrightarrow
\boldsymbol R_i,\xi_\alpha$:
\begin{eqnarray}
Z[\boldsymbol J,\tilde j]&=&
\left[
\prod_{i=1}^N\prod_{\alpha=1}^M
\lim_{\tilde D_\alpha\to  0}
\lim_{\eta_\alpha\to 0}
\int_{\boldsymbol R_i(0)=\boldsymbol R_{0,i}}^{\boldsymbol R_i(f)=\boldsymbol R_{f,i}}
{\cal D}\boldsymbol R_i{\cal  D}\xi_\alpha 
\right] 
Jac\exp
\left\{
-\sum_{i=1}^N\int_0^{t_f}dt
\textstyle\frac{(\dot{\boldsymbol R}_i-\boldsymbol f)^2}{4D}
\right\}
\nonumber\\
&\!\!\!\!\!\!\!\!\!\!\!\!\!\!\!\!\!\!\!\!\!\!\!\!\!\!\!\!\!\!\!\!\!\!\!\!\!\!
\!\!\!\!\!\!
\times&\!\!\!\!\!\!\!\!\!\!\!\!\!\!\!\!\!\!\!\!\!\!\!\!
\exp\left\{
-\sum_{\alpha=1}^M\int_0^{t_f}\textstyle\frac{(\eta\dot{\xi}_\alpha+
C_\alpha(\{ \boldsymbol  R\}))^2}{4\tilde  D_\alpha}  
\right\}
\exp\left\{
-\int_0^{t_f}dt\left[
\sum_{\alpha=1}^M\xi_\alpha\tilde j_\alpha+\imath\sum_{i=2}^N\boldsymbol
R_i\cdot(\boldsymbol J_{i} -\boldsymbol J_{i-1})
\right]\right\}\label{genfunbis}
\end{eqnarray}
where $Jac$ is the determinant of the transformation. In
principle, $Jac$ is the determinant of a complicated block
matrix:
\begin{equation}
Jac=\det\left|
\begin{array}{cc}
\frac{\delta\nu_{i,a}}{\delta R_{j,b}}&
\frac{\delta\nu_{i,a}}{\delta \xi_{\beta}}\\
\frac{\delta\tilde\nu_{\alpha}}{\delta R_{j,b}}&
\frac{\delta\tilde\nu_{\alpha}}{\delta \xi_{\beta}}
\end{array}
\right|\label{jacobianfull}
\end{equation}
The determinant in the right hand side of Eq.~(\ref{jacobianfull})
can however be simplified by noticing that
Eqs.~(\ref{Langevinequationsbis}) establish relationships only between
the $\boldsymbol\nu_i$'s and the $\boldsymbol R_j$'s. As a
consequence, it turns out that $\frac{\delta\nu_{i,a}}{\delta
  \xi_{\beta}}=0$
for $i=1,\ldots,N$, $a=1,2,3$ and $\beta=1,\ldots,M$. For this reason,
$Jac$ reduces to a product of two determinants:
\begin{equation}
Jac=\det\left|\frac{\delta\nu_{i,a}}{\delta R_{j,b}}\right|
\det\left|\frac{\delta\tilde\nu_{\alpha}}{\delta
  \xi_{\beta}}\right|\label{jacobiansimplification} 
\end{equation}
A simple way to prove the above identity is to eliminate first the
$\boldsymbol \nu_i$'s in the functional integral
(\ref{generatingfunctional}) using
Eqs.~(\ref{Langevinequationsbis}). Clearly, 
from those equations it turns out that
$\boldsymbol\nu_i(\boldsymbol R)=\dot{\boldsymbol
  R}_i-\boldsymbol f_i$. The jacobian of this transformation is
$\det\left|\frac{\delta\nu_{i,a}}{\delta R_{j,b}}\right|$. After
substituting everywhere in (\ref{generatingfunctional}) the
$\boldsymbol\nu_i$'s with their new expression in terms of the
$\boldsymbol R_i$'s, we may eliminate also the auxiliary noises
$\tilde \nu_\alpha$ exploiting Eqs.~(\ref{Langevinconstraints}).
As a result, the $\tilde\nu_\alpha$ become functions of the new
variables $\boldsymbol R_i$'s and $\xi_\alpha$'s. The jacobian
determinant of this second transformation is
$\det\left|\frac{\delta\tilde\nu_{\alpha}}{\delta 
  \xi_{\beta}}\right|$. This completes our proof. Alternatively,
Eq.~(\ref{jacobiansimplification}) may be easily checked by noticing
that in the calculation of the full determinant in the right hand side
of Eq.~(\ref{jacobianfull}) each element of the matrix
$\frac{\delta\tilde\nu_{\alpha}}{\delta R_{j,b}}$ is necessarily
multiplied by an element of the matrix $\frac{\delta\nu_{i,a}}{\delta
  \xi_{\beta}}$ which is equal to zero.

Explicitly, the determinants in Eq.~(\ref{jacobiansimplification})
read as follows:
\begin{equation}
\det\left|\frac{\delta\nu_{i,a}(t)}{\delta R_{j,b}(t')}\right|=
\det\left|
\left(
\delta_{ij}\delta_{ab}\frac{\partial}{\partial t}-\frac{
\partial f_{i,a}(t)
}{\partial R_{j,b}(t)}
\right)\delta(t-t')
\right|\label{qt1}
\end{equation}
with $f_{i,a}(t)=f_{i,a}(\boldsymbol R_1(t),\ldots,\boldsymbol
R_N(t))$ and
\begin{equation}
\det\left|\frac{\delta\tilde\nu_{\alpha}(t)}{\delta
  \xi_{\beta}(t')}\right|=\eta_\alpha\frac{\partial}{\partial
  t}\delta(t-t') \label{qt2}
\end{equation}
The only relevant determinant is that in Eq.~(\ref{qt1}). Determinants
of this type can be evaluated in closed form. We give
here only the result of the calculation referring the interested
reader for all details to standard books in this subject like for instance
 Ref.~\cite{ZinnJustin}:
\begin{equation}
\det\left|
\left(
\delta_{ij}\delta_{ab}\frac{\partial}{\partial t}-\frac{
\partial f_{i,a}(t)
}{\partial R_{j,b}(t)}
\right)\delta(t-t')
\right|
=\exp\left[
-\frac 14\int_0^{t_f}dt\sum_{i=1}^N\sum_{a=1}^3\frac{
\partial f_{i,a}(t)
}{\partial R_{i,a}(t)}
\right ]
\label{qt1evaluated}
\end{equation}
Putting the expression of the jacobian just computed in
Eq.~(\ref{genfunbis}) and taking the limit 
$\eta_\alpha\to0$, we obtain:
\begin{eqnarray}
Z[\boldsymbol J]&=&\left[\prod_{i=1}^N
\int_{\boldsymbol R_i(0)=\boldsymbol R_{0,i}}^{\boldsymbol R_i(f)=\boldsymbol R_{f,i}}
{\cal D}\boldsymbol R_i
\right]\exp\left\{-\frac 1{4D}\sum_{i=1}^N\int_0^{t_f}dt(\dot{\boldsymbol
  R}_i-\boldsymbol f_i)^2\right\}\nonumber\\
&\times&\exp\left[
-\frac 14\int_0^{t_f}dt\sum_{i=1}^N\sum_{a=1}^3\frac{
\partial f_{i,a}(t)
}{\partial R_{i,a}(t)}
\right ]
\exp\left\{
-\int_0^{t_f}dt\left[
\imath\sum_{i=2}^N\boldsymbol
R_i\cdot(\boldsymbol J_{i} -\boldsymbol J_{i-1})
\right]\right\}\nonumber\\
&\times&\left[\prod_{\alpha=1}^M
\lim_{\tilde D_\alpha\to 0}
\right]\exp\left[-\sum_{\alpha=1}^M\frac1{4\tilde D_\alpha}
\int_0^{t_f}dt C_\alpha^2(\{\boldsymbol R\})
\right]\label{genfuntris}
\end{eqnarray}
The $\xi_\alpha$'s decouple from the other degrees of freedom   in the limit
$\eta_\alpha\to 0$ and have been already integrated out in
Eq.~(\ref{genfuntris}). It remains to perform the
limit $\tilde D_\alpha\to 0$. To this purpose it is possible to apply
the formula 
\cite{FePaVi,FePy}:
\begin{equation}
\lim_{\sigma\to 0}e^{-\frac 1\sigma\int_0^{t_f}f^2(\{\boldsymbol
  R\})}=\delta(f(\{\boldsymbol R\})) 
\end{equation}
As a result, Eq.~(\ref{genfuntris}) becomes:
\begin{eqnarray}
Z[\boldsymbol J]&=&\left[\prod_{i=1}^N
\int_{\boldsymbol R_i(0)=\boldsymbol R_{0,i}}^{\boldsymbol
  R_i(f)=\boldsymbol R_{f,i}} {\cal D}\boldsymbol R_i
\right]\exp\left\{-\frac 1{4D}\sum_{i=1}^N\int_0^{t_f}dt(\dot{\boldsymbol
  R}_i^2+\boldsymbol f_i^2)\right\}\nonumber\\
&\times&\exp\left[
-\frac 14\int_0^{t_f}dt\sum_{i=1}^N\sum_{a=1}^3\frac{
\partial f_{i,a}(t)
}{\partial R_{i,a}(t)}
\right ]
\exp\left\{
-\int_0^{t_f}dt\left[
\imath\sum_{i=2}^N\boldsymbol
R_i\cdot(\boldsymbol J_{i} -\boldsymbol J_{i-1})
\right]\right\}\nonumber\\
&\times&\left[\prod_{\alpha=1}^M
\delta(C_\alpha(\{\boldsymbol R\}))
\right]\label{genfuncfinal}
\end{eqnarray}
This is the desired expression of the generating functional
$Z[\boldsymbol J]$.
In writing the above formula we have have assumed that the forces
$\boldsymbol f_i$ are conservative, so that it is possible to use the identity
$\int_0^{t_f}dt(\dot{\boldsymbol R}_i-\boldsymbol f_i)^2=
\int_0^{t_f}(\dot{\boldsymbol
  R}_i^2+\boldsymbol f_i^2)$ which is valid up to a term which is a
total derivative. 
\section{The example of an inextensible chain}
As an example of the method for imposing the constraints explained in
the previous Section, we consider here the generating functional
$Z^{i.c.}[\boldsymbol J]$ describing the dynamics of an inextensible
chain fluctuating in a viscous media. The chain is a discrete
mechanical system consisting of $N$ beads of mass $m$ connected
together by $N-1$ massless segments of fixed length $a$. The total
length of the chain is $L=Na$ and its total mass is $M=Nm$. For this
particular example the constraints (\ref{constraints}) take the form:
\begin{equation}
C_\alpha(\{\boldsymbol R\})=|\boldsymbol R_\alpha-\boldsymbol
R_{\alpha-1}|^2-a^2=0\qquad \qquad\alpha=2,\ldots,N
\end{equation}
The direct application of Eq.~(\ref{genfuncfinal}) to the inextensible
chain gives as a result the generating functional:
\begin{eqnarray}
Z^{i.c.}[\boldsymbol J]&=&\left[\prod_{i=1}^N
\int_{\boldsymbol R_i(0)=\boldsymbol R_{0,i}}^{\boldsymbol
  R_i(f)=\boldsymbol R_{f,i}} {\cal D}\boldsymbol R_i
\right]\exp\left\{-\frac 1{4D}\sum_{i=1}^N\int_0^{t_f}dt(\dot{\boldsymbol
  R}_i^2-\boldsymbol f_i^2)\right\}\nonumber\\
&\times&\exp\left[
-\frac 14\int_0^{t_f}dt\sum_{i=1}^N\sum_{a=1}^3\frac{
\partial f_{i,a}(t)
}{\partial R_{i,a}(t)}
\right ]
\exp\left\{
-\int_0^{t_f}dt\left[
\imath\sum_{i=2}^N\boldsymbol
R_i\cdot(\boldsymbol J_{i} -\boldsymbol J_{i-1})
\right]\right\}\nonumber\\
&\times&\left[\prod_{\alpha=2}^N
\delta(|\boldsymbol R_\alpha-\boldsymbol R_{\alpha-1}|^2-a^2)
\right]\label{genfuncinextchain}
\end{eqnarray}
In the following all external forces will be switched off,
i.~e. $\boldsymbol f_i=0$.
We will now consider the continuous limit of $Z^{i.c.}[\boldsymbol J]$:
\begin{equation}
a\to0\qquad\qquad N\to\infty\qquad\qquad Na=L=\mbox{const.}\label{scontlimit}
\end{equation}
This limit can be performed using the prescription of
Ref.~\cite{FePaVi}. 
The result is the partition function
\begin{equation}
{\cal Z}[\boldsymbol
  J]=\lim_{a\to0,N\to\infty,Na=L}Z^{i.c.}[\boldsymbol J]
\end{equation}
of what has been called the generalized nonlinear sigma model in
Ref.~\cite{FePaVi}:
\begin{equation}
{\cal Z}[\boldsymbol J]=\int_{\boldsymbol R(0,s)=\boldsymbol
  R_0(s)}^{\boldsymbol R(t_f,s)=\boldsymbol R_f(s)} {\cal
  D}\boldsymbol
R(t,s)
e^{-\int_0^{t_f}dt\int_0^L\left(c\dot{\boldsymbol
    R}^2(t,s)+\boldsymbol R(t,s)\cdot\boldsymbol J'(t,s)
  \right )}\delta\left(|\boldsymbol R'(t,s)|^2-1\right)\label{genfuncgnlsm}
\end{equation}
In the above equation 
$\boldsymbol R_0(s)$ and $\boldsymbol R_f(s)$ represent the continuous
version of the discrete initial and final conformations of the chain
given in Eq.~(\ref{boundaryconditions}). Moreover,
$s$ is the arc-length on the chain and
$\boldsymbol R'(t,s) =\frac{\partial R(t,s)}{\partial s}$. Finally,
\begin{equation}
c=\frac{1}{2k_BT\tau}\frac M{2L}\label{cdef}
\end{equation}
with $k_B$ being the Boltzmann constant. 
The parameter $c$ contains physical constants like the temperature
of the viscous medium $T$ and the relaxation time of the infinitesimal
beads $\tau$.   In the frame of the
duality between statistical physics and quantum mechanics, the
quantity $2k_BT\tau$ is the analog of the Planck constant \cite{rice}. It
describes the uncertainties in determining the positions and the
momenta of the beads because of the collisions with the molecules of
the surrounding medium.

Up to now, the generating functional ${\cal Z}[\boldsymbol J]$ has been
computed in the semiclassical approximation \cite{FePaVi} or it has
been linearized 
using a variational method \cite{FePy}.
In the rest of this Section an attempt to go beyond these
approximations will be presented. 
In order to avoid as much as possible complications with the boundary
conditions, we will suppose that the chain is ring-shaped. Accordingly,
throughout the rest of this Section all the fields will satisfy
periodic boundary conditions of the type:
\begin{equation}
\boldsymbol R(t,s+L)=\boldsymbol R(t,s)\label{boundcondbigr}
\end{equation}
The functional Dirac delta function
imposing the constraints in Eq.~(\ref{genfuncgnlsm}) may be treated by
introducing the new field $\boldsymbol r(t,s)$ and rewriting  ${\cal
  Z}[\boldsymbol J]$ as follows:
\begin{eqnarray}
{\cal Z}[\boldsymbol J]&=&\int_{\boldsymbol R(0,s)=\boldsymbol
  R_0(s)}^{\boldsymbol R(t_f,s)=\boldsymbol R_f(s)} 
{\cal
  D}\boldsymbol
R(t,s)
\int_{\boldsymbol r(0,s)=\boldsymbol R'_0(s)}^{\boldsymbol r(t_f,s)=\boldsymbol R'_f(s)}
 {\cal
  D}\boldsymbol r(t,s)
e^{-\int_0^{t_f}dt\int_0^L\left(c\dot{\boldsymbol
    R}^2(t,s)+\boldsymbol R(t,s)\cdot\boldsymbol J'(t,s)
  \right )}
\nonumber\\
&\times&\delta(\boldsymbol r(t,s)-\boldsymbol
R'(t,s))\delta\left(\boldsymbol 
r^2(t,s)-1\right) \label{zjinitial}
\end{eqnarray}
When $t=0$
and $t=t_f$, $\boldsymbol r(t,s)$ obeys respectively the boundary
conditions: 
\begin{equation}
\boldsymbol r(0,s)=\boldsymbol R'_i(s)\qquad\mbox{and}
\qquad \boldsymbol r(t_f,s)=\boldsymbol R'_f(s)\label{bcrsmall}
\end{equation}
At this point we express the first
delta function appearing in the right hand side of
Eq.~(\ref{zjinitial}) using the Fourier representation: 
\begin{equation}
\delta(\boldsymbol r-\boldsymbol
R')=\int{\cal D}\boldsymbol ke^{\imath\int_0^{t_f}dt\int_0^Lds\boldsymbol
  k\cdot(\boldsymbol r-\boldsymbol R')}\label{ftdeltarrprime}
\end{equation}
The new Fourier variable $\boldsymbol k=\boldsymbol k(t,s)$ must be free to
vary in the interval $(-\infty,+\infty)$ over the whole domain
$[0,L]\times(0,t_f)$ of $s$ and $t$ in order to impose the relation
$\boldsymbol r(t,s)=\boldsymbol R'(t,s)$. When $t=0$ and
$t=t_f$, instead, the values of $\boldsymbol r(0,s)$ and
$\boldsymbol r(t_f,s)$ have been already fixed in the generating
functional of Eq.~(\ref{zjinitial})
to be those of
Eq.~(\ref{bcrsmall}). For this reason, at the instants $t=0$ and
$t=t_f$ we may chose
for $\boldsymbol k(t,s)$  Dirichlet boundary
conditions, i.~e.
\begin{equation}
\boldsymbol k(0,s)=\boldsymbol k(t_f,s)=0\label{dirichletbcfork}
\end{equation}
Applying
Eq.~(\ref{ftdeltarrprime}) in 
Eq.~(\ref{zjinitial}), it is possible to split ${\cal Z}[\boldsymbol
  J]$
into a path integral over $\boldsymbol k(t,s)$ of the product of two different
contributions:
\begin{equation}
{\cal Z}[\boldsymbol J]=\int{\cal D}\boldsymbol k{\cal
  Z}_1[\boldsymbol k,\boldsymbol J]{\cal
  Z}_2[\boldsymbol k]\label{disentanglement}
\end{equation}
where
\begin{equation}
{\cal Z}_1[\boldsymbol k,\boldsymbol J]=\int_{\boldsymbol
  R(0,s)=\boldsymbol R_0(s)}^{\boldsymbol R(t_f,s)=\boldsymbol
  R_f(s)}{\cal D}\boldsymbol R \exp\left\{-\int_0^{t_f}dt\int_0^Lds \left[
    c\dot{\boldsymbol R}^2+\imath(\boldsymbol J'+\boldsymbol
    k')\cdot\boldsymbol R\right]\right\}\label{z1kJ}
\end{equation}
and
\begin{equation}
{\cal Z}_2[\boldsymbol k]=\int_{\boldsymbol r(0,s) =\boldsymbol
  R_0'(s)}^{\boldsymbol r(t_f,s) =\boldsymbol
  R_f'(s)}{\cal D}\boldsymbol r 
e^{+\imath\int_0^{t_f}dt\int_0^Lds\boldsymbol 
k\cdot\boldsymbol r} \delta(\boldsymbol r^2-1)\label{z2k}
\end{equation}
As we see, the integration over $\boldsymbol R(t,s)$ 
is completely disentangled from that over $\boldsymbol r(t,s)$, but
both fields 
interact with  $\boldsymbol k(t,s)$.

First, the functional ${\cal Z}_1[\boldsymbol k,\boldsymbol J]$ will
be computed. To this purpose, it is convenient to perform the
change of variables:
\begin{equation}
\boldsymbol R=\boldsymbol R_{cl}+\delta \boldsymbol R
\end{equation}
Here $\boldsymbol R_{cl}$ is a solution of the free classical equation of
motion $\ddot{\boldsymbol R}_{cl}=0$ and satisfies the same boundary
conditions as $\boldsymbol R$. Consequently, the fluctuation $\delta
\boldsymbol R$ obeys the Dirichlet boundary conditions:
\begin{equation}
\delta\boldsymbol R(t_f,s)=\delta\boldsymbol R(0,s)=0
\end{equation}
Moreover, both $\boldsymbol R_{cl}(t,s)$ and $\delta\boldsymbol R$ are
periodic in $s$.
After a gaussian integration over $\delta \boldsymbol R$, we
obtain:
\begin{equation}
{\cal Z}_1[\boldsymbol k,\boldsymbol J]=e^{-S_{cl}(\boldsymbol
  k,\boldsymbol J)}\exp\left\{
-\frac 14\int_0^{t_f}dtdt'\int_0^LdsG(t,t')(\boldsymbol
J'(t,s)+\boldsymbol k'(t,s))\cdot(\boldsymbol
J'(t',s)+\boldsymbol k'(t',s))
\right\}\label{z1kjfinal}
\end{equation}
where
\begin{equation}
S_{cl}(\boldsymbol
  k,\boldsymbol J)=\int_0^{t_f}dt\int_0^Lds\left[
c\dot{\boldsymbol R}^2_{cl}+\imath(\boldsymbol J'+\boldsymbol
k')\cdot\boldsymbol R_{cl}
\right]\label{classicalsector}
\end{equation}
while $G(t,t')$ is the Green function satisfying the equation:
\begin{equation}
2c\frac{\partial^2G(t,t')}{\partial t^2}=\delta(t-t')
\end{equation}
and Dirichlet boundary conditions.
Explicitly:
\begin{equation}
G(t,t')=\frac1{2ct_f}
\left[
\frac{(t+t')}{2}-\frac{|t-t'|}{2}
\right]
\left[
\frac{(t+t')}{2}+\frac{|t-t'|}{2}
-t_f\right]
\end{equation}
Next, we consider the functional ${\cal Z}_2[\boldsymbol k]$. 
This quantity closely resembles the Fourier transform of the
probability distribution of a freely jointed chain appearing in the
statistical 
mechanics of polymers. It may be
computed exactly in a similar way using a discretization procedure.
We have already seen how to pass from the continuous interval $[0,L]$ to
its discrete approximation in the case of the arc-length $s$, see
e.~g. Eq.~(\ref{scontlimit}). Analogously, we discretize the time
interval $[0,t_f]$ by replacing it with a lattice of ${\cal M}$ sites
and link length $b$. The continuous limit is performed by requiring
that:
\begin{equation}
b\to0\qquad\qquad{\cal M}\to+\infty\qquad\qquad{\cal M}b=t_f
\end{equation}
Ignoring for the moment the nontrivial time boundary conditions of
$\boldsymbol r(t,s)$, a
straightforward calculation shows that ${\cal Z}_2[\boldsymbol k]$
is given by:
\begin{equation}
{\cal Z}_2[\boldsymbol k]=\lim_{\substack
  {a\to0,N\to+\infty,Na=L\\b\to0,{\cal 
      M}\to+\infty,{\cal M}b=t_f
}}\prod_{n=1}^N\prod_{m=1}^{\cal M}\frac{\sin(ab|\boldsymbol
  k_{nm}|)}{ab|\boldsymbol k_{nm}|} \label{discretizedexact2}
\end{equation}
Here $ \boldsymbol k_{nm}$ is a shorthand notation for $\boldsymbol
k(t_m,s_n)$. 
Apart from the boundary conditions, the problem with
Eq.~(\ref{discretizedexact2}) is the evaluation of 
the continuous limit of its right hand side.  This is a nontrivial
task.
The naive prescription
used in the statistical mechanics of polymers
\cite{doiedwards,kleinertpath}, i.~e.
\begin{equation}
\frac{\sin(ab|\boldsymbol  k_{nm}|)}
{ab|\boldsymbol k_{nm}|}
\sim 1-(ab|\boldsymbol
  k_{nm}|)^2\sim
e^{-{(ab|\boldsymbol k_{nm}|)^2}}
\end{equation}
fails to provide the correct result.
For that reason, in order to evaluate ${\cal Z}_2[\boldsymbol k]$, we
 expand the exponential
$e^{\imath \int_0^{t_f}dt\int_0^Lds\boldsymbol k \cdot\boldsymbol r}$
in power series:
\begin{equation}
{\cal Z}_2[\boldsymbol
  k]=\sum_{\sigma=0}^{+\infty}\frac{(-\imath)^{2\sigma}}{2\sigma!}
I_{2\sigma}\label{z2powers} 
\end{equation}
where
\begin{equation}
I_{2\sigma}=\int_{\boldsymbol r(0,s)=\boldsymbol R'_0(s)}^{\boldsymbol
  r(t_f,s)=\boldsymbol R'_f(s)} {\cal D}\boldsymbol
r(t,s)\delta(\boldsymbol 
r^2(t,s)-1)
\prod_{l=1}^{2\sigma}\int_0^{t_f}dt_l\int_0^Lds_lk_{a_l}(t_l,s_l)r_{a_l}(t_l,s_l)
\end{equation}
In the above equation the sum over repeated spatial indexes
$a_1,\ldots,a_{2\sigma}=1,2,3$ is understood.
Due to the presence of the functional delta function, it is difficult
to apply the Wick theorem in order to integrate over the field
$\boldsymbol r(t,s)$. However, a
strategy to evaluate $I_{2\sigma}$  becomes clear
after the
discretization of both variables $t$ and $s$. Calling
$I_{2\sigma}\{N,{\cal M},a,b\}$ the 
discretized version of $I_{2\sigma}$, we obtain:
\begin{eqnarray}
I_{2\sigma}\{N,{\cal M},a,b\}&=&\prod_{n=1}^N\prod_{m=2}^{{\cal M}-1}\int
d\boldsymbol r_{nm}\delta(\boldsymbol
r_{nm}^2-1)\prod_{l=1}^{2\sigma}\left(
\sum_{n_l=1}^{N}a\sum_{m_l=2}^{{\cal M}-1}b\,\,k_{a_l,n_lm_l}r_{a_l,n_lm_l}
\right)\nonumber\\
&\times&
\prod_{n=1}^N\int d\boldsymbol r_{n1}
 d\boldsymbol r_{n{\cal M}}
{A}_1{A}_{\cal M}\label{I2sigmadiscrvers}
\end{eqnarray}
with the boundary conditions at the initial and final times fixed by
Dirac delta functions inside the factors ${A}_1$ and ${A}_{\cal M}$:
\begin{equation}
A_1=\prod_{n=2}^N\delta\left(\boldsymbol r_{n1}-\frac{(\boldsymbol
  R_{0,n}-\boldsymbol R_{0,n-1})}{a}\right)\qquad\mbox{and}\qquad
A_{\cal M}=\prod_{n=2}^N\delta\left(\boldsymbol r_{n{\cal M}}-\frac{(\boldsymbol
  R_{f,n}-\boldsymbol R_{f,n-1})}{a}\right)
\end{equation}
Let us note that, due to the choice of Dirichlet boundary condition
for $\boldsymbol k(t,s)$, the boundary values of this field are all
zero, i.~e. $\boldsymbol k_{n1}=\boldsymbol k_{n{\cal M}}=0$ for
$n=1,\ldots,N$. 
For this reason, the sums over the indices $m_l$ in
Eq.~(\ref{I2sigmadiscrvers}) have been restricted in the range $2\le
m_l\le {\cal M}-1$ for $l=1,\ldots,2\sigma$. As a consequence, 
the variables at the boundary $\boldsymbol r_{n1}$ and
$\boldsymbol r_{n{\cal M}}$ in Eq.~(\ref{I2sigmadiscrvers}) are not
present in the product $\prod_{l=1}^{2\sigma}\left(
\sum_{n_l=1}^{N}a\sum_{m_l=2}^{{\cal M}-1}b\,\,k_{a_l,n_lm_l}r_{a_l,n_lm_l}
\right)$.
They enter only in the terms $A_1$ and $A_{\cal M}$ and are not
mixed with the other degrees of freedom to be integrated. Thanks
to this fact, the 
integrals over $\boldsymbol r_{n1}$ and
$\boldsymbol r_{n{\cal M}}$ for $n=1,\ldots,N$ can be easily factored
out from the expression of  $I_{2\sigma}\{N,{\cal M},a,b\}$.
The calculation of $I_{2\sigma}\{N,{\cal M},a,b\}$ requires the
integration over the remaining $3N({\cal M}-2)$ variables $\boldsymbol
r_{nm}$ for $1\le n\le N$ and $2\le m\le {\cal M}-1$ with 
the measure $d\boldsymbol r_{nm}\delta(\boldsymbol r_{nm}^2-1)$. The
integrand consists in a sum of terms in which $2\sigma$ 
components $r_{a_l,n_lm_l}$ of the vectors $\boldsymbol
r_{n_lm_l}$ with various combinations
of the indexes $a_l,m_l,n_l$
are multiplied together. Each of such components may appear inside the
product with powers of any order comprised between $0$ and
$2\sigma$.
Clearly, only the integration over terms in which 
the indexes $a_l$ of the components and the discrete indexes $n_l,m_l$
are present in even combinations will  not vanish
identically. Basing
ourselves on these observation, it is possible to establish the following
rules in order to evaluate $I_{2\sigma}\{N,{\cal
  M},a,b\}$:
\begin{enumerate}
\item 
Rearrange the terms appearing in $I_{2\sigma}\{N,{\cal
  M},a,b\}$ in such a way that the products of the $2\sigma$
  components $r_{a_l,n_lm_l}$'s will be grouped into products of
  $2l-$plets for $l=1,\ldots,\sigma$ characterized by the fact that
  inside each multiplet all the components have the same indices
  $n_l$ and $m_l$.\label{item1}
\item Eliminate the spatial indices $a_l$ by using the formula
\begin{equation}
\sum_{i_l=1}^3k_{a_l,n_lm_l}r_{a_l,n_lm_l}=|\boldsymbol
k_{n_lm_l}||\boldsymbol r_{n_lm_l}| \cos\theta_{n_lm_l}
\end{equation}\label{item2}
\item Use the fact that, due to the delta functions
  $\delta(\boldsymbol r^2_{nm}-1)$ it is possible to put $|\boldsymbol
  r_{n_lm_l}|=1$ and integrate over the angles $\theta_{n_lm_l}$
  using the formula:
\begin{equation}
\int_0^\pi
d\theta_{mn}\sin\theta_{nm}\cos^{2i}\theta_{nm}=2\frac1{2i+1}\qquad\qquad
i=0,1,\ldots 
\end{equation}
\label{item3} 
\end{enumerate}
Let $K_l$ be the number of $2l-$plets mentioned in
\ref{item1}. Clearly, the possible values of $K_l$ are limited by the
condition  $\sum_{l=1}^\sigma 2lK_l=2\sigma$. This implies for
instance that $K_l=1$ for $l>\sigma$. Moreover, the number of
combinations for dividing $2\sigma$ objects in a number $K_1$ of
pairs, $K_2$ of $4-$plets etc. is given by:
\begin{equation}
K=\frac{2\sigma!}{2K_1!4K_2!\cdots 2(\sigma-1)K_{\sigma-1}!}
\end{equation}
Following steps \ref{item1}. and \ref{item2}. of the above
prescription, we may write:
\begin{eqnarray}
I_{2\sigma}\{N,{\cal
  M},a,b\}&=&\sum_{\substack{K_1,\ldots,K_\sigma\\ \sum_{l=1}^\sigma2lK_l=2\sigma}} 
\frac{2\sigma!}{2K_1!4K_2!\cdots 2(\sigma-1)K_{\sigma-1}!}
\prod_{l=1}^\sigma\sum_{n_{l,1},\ldots,n_{l,K_l}=1}^N
\!\!\!\!\!\!a^{K_l}
\sum_{m_{l,1},\ldots,m_{l,K_l}=1}^{\cal
  M}\!\!\!\!\!\!b^{K_l}\nonumber\\
&\!\!\!\!\!\!\!\!\!\!\!\!\!\!\!\!\!\!\!\!\!\!\!\!\!\!\!\!\!\!\!\!\!\!\!\!\!\!\!\!\!\!\!\!\!\!\!\!\!\!\!\!\!\!\!\!\!\!\!\!\!\!\!\!\!\!\!\!\!\!\!\!\!\!\!\!\!\!\!\!\!\!\!\!
\times&
\!\!\!\!\!\!\!\!\!\!\!\!\!\!\!\!\!\!\!\!\!\!\!\!\!\!\!\!\!\!\!\!\!\!\!\!\!\!\!\!\!\!
|\boldsymbol k_{n_{l,1}m_{l,1}}|^{2l}
\ldots |\boldsymbol k_{n_{l,K_l}m_{l,K_l}}|^{2l}\prod_{n,m=1}^{N,{\cal
M}} \int_0^{+\infty}d|\boldsymbol r_{nm}|\delta(|\boldsymbol
r_{nm}|^2-1)|\boldsymbol r_{n_{l,1}m_{l,1}}|^{2l}
\ldots |\boldsymbol r_{n_{l,K_l}m_{l,K_l}}|^{2l}\nonumber \\
&\!\!\!\!\!\!\!\!\!\!\!\!\!\!\!\!\!\!\!\!\!\!\!\!\!\!\!\!\!\!\!\!\!\!\!\!\!\!\!\!\!\!\!\!\!\!\!\!\!\!\!\!\!\!\!\!\!\!\!\!\!\!\!\!\!\!\!\!\!\!\!\!\!\!\!\!\!\!\!\!\!\!\!\!
\times&
\!\!\!\!\!\!\!\!\!\!\!\!\!\!\!\!\!\!\!\!\!\!\!\!\!\!\!\!\!\!\!\!\!\!\!\!\!\!\!\!\!\!
\int_0^{2\pi}d\phi_{nm}\int_0^\pi
d\theta_{nm}\sin\theta_{nm}\cos^{2l}\theta_{n_{l,1}m_{l,1}} \ldots
\cos^{2l}\theta_{n_{l,K_l}m_{l,K_l}} A_1 A_{\cal M}
\end{eqnarray}
It is now possible to perform the remaining integrations according to
prescription \ref{item3}. The result is:
\begin{eqnarray}
I_{2\sigma}\{N,{\cal
  M},a,b\}&=&Z_2[0;N,{\cal
    M},a,b]\sum_{\substack{K_1,\ldots,K_\sigma\\ \sum_{l=1}^\sigma2lK_l=2\sigma}}  
\frac{2\sigma!}{2K_1!4K_2!\cdots 2(\sigma-1)K_{\sigma-1}!}
\nonumber\\
&\!\!\!\!\!\!\!\!\!\!\!\!\!\!\!\!\!\!\!\!\!\!\!\!\!\!\!\!\!\!\!\!\!\!\!\!\!\!\!\!\!\!\!\!\!\!\!\!\!\!\!\!\!\!\!\!\!\!\!\!\!\!\!\!\!\!\!\!\!\!\!\!\!\!\!\!\!\!\!\!\!\!\!\!
\times&
\!\!\!\!\!\!\!\!\!\!\!\!\!\!\!\!\!\!\!\!\!\!\!\!\!\!\!\!\!\!\!\!\!\!\!\!\!\!\!\!\!\!
\prod_{l=1}^\sigma\sum_{n_{l,1},\ldots,n_{l,K_l}=1}^N
\!\!\!\!\!\!a^{K_l}
\sum_{m_{l,1},\ldots,m_{l,K_l}=1}^{\cal
  M}\!\!\!\!\!\!b^{K_l}
|\boldsymbol k_{n_{l,1}m_{l,1}}|^{2l}
\ldots |\boldsymbol k_{n_{l,K_l}m_{l,K_l}}|^{2l}\frac 1
       {(2l+1)^{K_l}}\label{Isigmafinaldiscrete} 
\end{eqnarray}
where
\begin{equation}
 Z_2[0;N,{\cal
    M},a,b]=\prod_{n,m=1}^{N,{\cal
M}} \int_0^{+\infty}d|\boldsymbol r_{nm}|\delta(|\boldsymbol
r_{nm}|^2-1)|\int_0^{2\pi}d\phi_{nm}\int_0^\pi
d\theta_{nm}\sin\theta_{nm} A_1 A_{\cal M}
\end{equation}
is the discrete version of ${\cal Z}_2[\boldsymbol k=0]$.
The continuous limit of $I_{2\sigma}\{N,{\cal M},a,b\}$ starting from
Eq.~(\ref{Isigmafinaldiscrete}) gives:
\begin{eqnarray}
I_{2\sigma}&=&{\cal Z}_2[0]\sum_{\substack{K_1,\ldots,K_\sigma\\ \sum_{l=1}^\sigma2lK_l=2\sigma}} 
\frac{2\sigma!}{2K_1!4K_2!\cdots 2(\sigma-1)K_{\sigma-1}!}
\nonumber\\
&\!\!\!\!\!\!\!\!\!\!\!\!\!\!\!\!\!\!\!\!\!\!\!\!\!\!\!\!\!\!\!\!\!\!\!\!\!\!\!\!\!\!\!\!\!\!\!\!\!\!\!\!\!\!\!\!\!\!\!\!\!\!\!\!\!\!\!\!\!\!\!\!\!\!\!\!\!\!\!\!\!\!\!\!\!\!\!\!\!\!\!\!
\times&
\!\!\!\!\!\!\!\!\!\!\!\!\!\!\!\!\!\!\!\!\!\!\!\!\!\!\!\!\!\!\!\!\!\!\!\!\!\!\!\!\!\!\!\!\!\!
\prod_{l=1}^\sigma
\frac 1 {(2l+1)^{K_l}}
\int_0^{t_f}dt_{l,1}\int_0^Lds_{l,1}\left|
\boldsymbol k(t_{l,1},s_{l,1})
\right|^{2l}\cdots\int_0^{t_f}dt_{l,K_l}\int_0^Lds_{l,K_l}\left|
\boldsymbol k(t_{l,K_l},s_{l,K_l})
\right|^{2l}\label{Isigmacontinuous}
\end{eqnarray}
with
\begin{equation}
{\cal Z}_2[0]=\lim_{\substack{
a\to0,N\to+\infty,Na=L\\
b\to0,{\cal M}\to+\infty,{\cal M}b=t_f
}} Z_2[0;N,{\cal M},a,b]
\end{equation}
${\cal Z}_2[0]$ contains the dependence on the
boundary values of the fields $\boldsymbol R(t,s)$.
Finally, substituting Eq.~(\ref{Isigmacontinuous}) in
Eq.~(\ref{z2powers}), we obtain the explicit expression of ${\cal
  Z}_2[\boldsymbol k]$.
Having derived the 
 functionals ${\cal Z}_1[\boldsymbol
  k,\boldsymbol J]$ and ${\cal Z}_2[\boldsymbol k]$, we may compute
 the full generating functional ${\cal Z}[\boldsymbol J]$
of Eq.~(\ref{disentanglement}):
\begin{eqnarray}
{\cal Z}[\boldsymbol J]&=&{\cal Z}_2[0]\int{\cal D}\boldsymbol k(t,s)
e^{-S_{cl}(\boldsymbol k,\boldsymbol J)} e^{
-\frac 14\int_0^{t_f}dtdt'\int_0^LdsG(t,t')(\boldsymbol
J'(t,s)+\boldsymbol k'(t,s) ) \cdot(\boldsymbol J'(t',s)+\boldsymbol
k'(t',s)) 
}\nonumber\\
&\times&\sum_{\sigma=0}^{+\infty}\frac{(-\imath)^{2\sigma}}{2\sigma!}
\sum_{\substack{K_1,\ldots,K_\sigma\\ \sum_{l=1}^\sigma2lK_l=2\sigma}} 
\frac{2\sigma!}{2K_1!4K_2!\cdots 2(\sigma-1)K_{\sigma-1}!}
\prod_{l=1}^\sigma\frac 1 {(2l+1)^{K_l}}\nonumber\\
&\times&\int_0^{t_f}dt_{l,1}\int_0^Lds_{l,1}\left|
\boldsymbol k(t_{l,1},s_{l,1})
\right|^{2l}\cdots\int_0^{t_f}dt_{l,K_l}\int_0^Lds_{l,K_l}\left|
\boldsymbol k(t_{l,K_l},s_{l,K_l})
\right|^{2l}\label{fullpertzj}
\end{eqnarray}
and perform the remaining integrations
over the fields $\boldsymbol k(t,s)$.
\begin{figure}
\centering
\includegraphics[width=10cm]{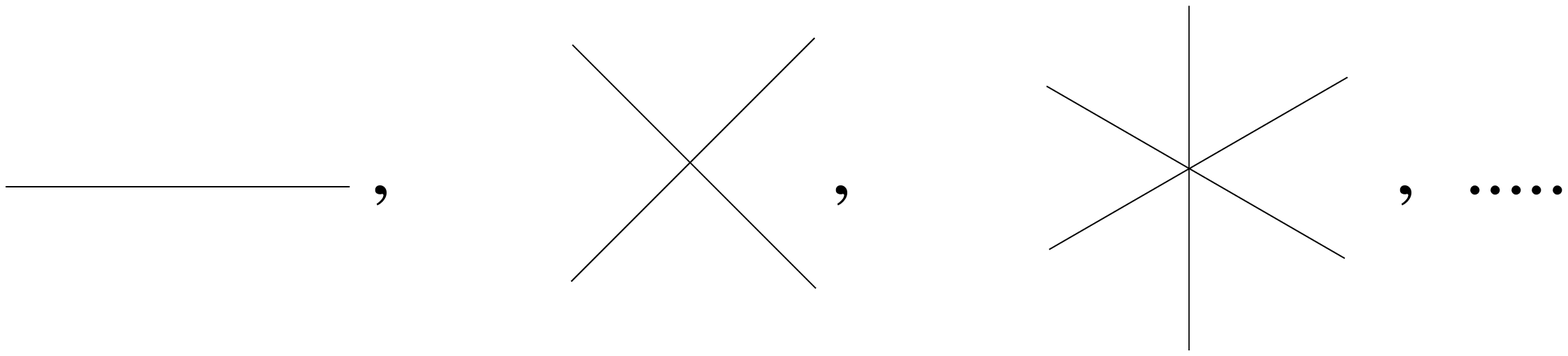}
\caption{This figure shows the Feynman diagrams corresponding to the
  terms $\int_0^{t_f}dt\int_0^Lds|\boldsymbol k(t,s)|^2 $,
$\int_0^{t_f}dt\int_0^Lds|\boldsymbol k(t,s)|^4 $,
$\int_0^{t_f}dt\int_0^Lds|\boldsymbol k(t,s)|^6 $,$\ldots$ appearing
  in Eq.~(\ref{Isigmacontinuous}).}\label{figone}
\end{figure}
Let us note that the action in Eq.~(\ref{fullpertzj}) is gaussian. The
classical sector present in the term $S_{cl}(\boldsymbol k,\boldsymbol
J)$ of Eq.~(\ref{classicalsector}) introduces a coupling of
$\boldsymbol k(t,s)$ with an external current which is proportional to
the classical conformation $\boldsymbol R_{cl}(t,s)$.
It is actually possible to rewrite the generating functional ${\cal
  Z}[\boldsymbol J]$ in a much more compact form by computing the
following functional:
\begin{eqnarray}
{\cal Z}[\boldsymbol J,\boldsymbol \mu]&=&
\exp\{-\int_0^{t_f}dt\int_0^Lds c\dot{\boldsymbol R}^2_{cl}(t,s)\}\nonumber\\
&\times&\exp\left\{
-\imath\int_0^{t_f}dtdt'\int_0^LdsG(t,t')(\boldsymbol
J'(t,s)+\boldsymbol k'(t,s))\cdot(\boldsymbol
J'(t',s)+\boldsymbol k'(t',s))
\right\}\nonumber\\
&\times&\exp\left\{-\imath\int_0^{t_f}dt\int_0^Lds\boldsymbol
  \mu(t,s)\cdot\boldsymbol k(t,s)\right \}\label{zfunctextchain}
\end{eqnarray}
where $\boldsymbol \mu(t,s)$ is the external current that is needed to
generate the correlation functions of the field $\boldsymbol k(t,s)$.
${\cal Z}[\boldsymbol J,\boldsymbol \mu]$ is the double generating
functional of the correlation functions of the physical fields
$\boldsymbol R(t,s)$ and of the auxiliary fields $\boldsymbol k(t,s)$.
Knowing ${\cal Z}[\boldsymbol J,\boldsymbol \mu]$, the
generating functional ${\cal Z}[\boldsymbol J]$ may be expanded as
follows:
\begin{eqnarray}
{\cal Z}[\boldsymbol
  J]&=&\sum_{\sigma=0}^{+\infty}
\sum_{\substack{K_1,\ldots,K_\sigma\\ \sum_{l=1}^\sigma2lK_l=2\sigma}}
\prod_{l=1}^\sigma\frac 1{(2lK_l)!}\frac 1{(2l+1)^{K_l}}\nonumber\\
&
\!\!\!\!\!\!\!\!\!\!\!\!\!\!\!\!
\times&
\!\!\!\!\!\!\!\!
\left(
\frac{\delta^2}{\delta\boldsymbol\mu(t_{l,1},s_{l,1})\cdot
  \delta\boldsymbol\mu(t_{l,1},s_{l,1})}
\right)^l\cdots
\left(\frac{\delta^2}
{\delta\boldsymbol\mu(t_{l,K_l},s_{l,K_l})\cdot
  \delta\boldsymbol\mu(t_{l,K_l},s_{l,K_l})}
\right)^l\left.{\cal Z}[\boldsymbol J,\boldsymbol \mu]\right|_{\boldsymbol\mu=0}
\label{pertexp}
\end{eqnarray}
${\cal Z}[\boldsymbol J,\boldsymbol\mu]$ can
be computed in closed form. Its expression is given below:
\begin{eqnarray}
{\cal Z}[\boldsymbol J,\boldsymbol\mu]&=&{\cal Z}_2[0]\exp\left\{
-\int_0^{t_f}dt\int_0^Lds\left(
c\dot{\boldsymbol R}_{cl}^2+\imath\boldsymbol J'\cdot\boldsymbol R_{cl}
-\imath \boldsymbol\mu\cdot\boldsymbol J\right)
\right\}\nonumber\\
&\times&\exp\left\{
-\int_0^{t_f}dtdt'\int_0^Ldsds'{\cal
  G}(t,s;t',s')\boldsymbol{\tilde\mu}(t,s)\cdot \boldsymbol{\tilde\mu}(t',s')
\right\}\label{extendedgenfun}
\end{eqnarray}
where
\begin{equation}
{\cal
  G}(t,s;t',s')=\frac c2\frac{\partial^2\delta(t-t')}{\partial
  t^{\prime 2}}\sum_{n=1}^{+\infty}\frac{L^2}{(2\pi
  n)^2}\sin\frac{2\pi ns}{L}\sin\frac{2\pi ns'}{L}\label{propagatofin}
\end{equation}
and
\begin{equation}
\tilde{\boldsymbol \mu}=\boldsymbol\mu+\boldsymbol R_{cl}
\end{equation}
It is straightforward to realize that at each  order in
the index $\sigma$, the
series 
in Eq.~(\ref{pertexp}) contains products of vertices with $2l$
external legs  
like those shown in Fig.~\ref{figone}
for
$l=0,1,2,\ldots, \sigma$. The
legs of the $2l-$vertices are contracted in all possible ways in
Eqs.~(\ref{fullpertzj}) or (\ref{pertexp}) using the propagator
(\ref{propagatofin}). 
Convergence is granted for small values of the constant $c$.
It is important to stress that the term ${\cal Z}_2[0]$, which
contains a nontrivial dependence on the boundary conformations of
$\boldsymbol R(t,s)$, appears in Eq.~(\ref{extendedgenfun}) as an
overall factor that can be easily eliminated by choosing the
normalization of the correlation functions.
\section{A brief digression on the nonlinear sigma model} 
In this Section we will apply the method illustrated in the case of
the inextensible chain to a two dimensional nonlinear sigma model with
action: 
\begin{equation}
S_{sm}[ J]=
\int_Md^2x\left[\frac
  g2\left(\partial_i\boldsymbol\phi\right)^2+\boldsymbol J\boldsymbol
  \phi\right] 
\end{equation}
where $\partial_i=\frac{\partial}{\partial x_i}$, $x=(x_1,x_2)$ and
$M$ is a two dimensional manifold with Euclidean signature. To avoid
complications with the choice of boundary conditions, $M$ is chosen to
be a torus, so that periodic boundary conditions should be implemented
for $x_1$ and $x_2$.
The vector field $\boldsymbol \phi(x)$ is subjected to the constraint:
\begin{equation}
\boldsymbol \phi^2(x)=1
\end{equation}
The generating functional of the nonlinear sigma model in path
integral form looks very similar to that of the inextensible chain:
\begin{equation}
Z_{sm}[J]=\int{\cal D}\boldsymbol\phi e^{-S_{sm}[J]}\delta(\boldsymbol
\phi^2-1)
\end{equation}
Introducing the new fields $\boldsymbol r(x)$ and $\boldsymbol k(x)$,
the integration over the physical field $\boldsymbol \phi$ can be
performed without worrying about the functional Dirac delta function
$\delta(\boldsymbol
\phi^2-1)$. The result of this integration is:
\begin{equation}
{\cal Z}_{sm}[J]=\int{\cal D}\boldsymbol k(x)e^{-\frac
  14\int_Md^2xd^2yG(x,y)(\boldsymbol J(x)+\boldsymbol k(x))\cdot
(\boldsymbol J(x)+\boldsymbol k(x))
}{\cal Z}_2[k]
\end{equation}
Apart from the different topology of the underlying manifold $M$,
${\cal Z}_2[k]$ is the same as the functional ${\cal Z}_2[\boldsymbol
  k]$ given in Eq.~(\ref{z2k}):
\begin{equation}
{\cal Z}_2[k]=\int{\cal D}\boldsymbol r(x)e^{\imath\int_M\boldsymbol
  k(x)\cdot\boldsymbol r(x)}\delta(\boldsymbol r^2(x)-1)
\end{equation}
One may thus evaluate this functional exactly as in the previous
Section. The result is an expansion of ${\cal Z}_2[k]$ at all orders
in powers of $\boldsymbol k(x)$ which converges for small values of
the coupling constant $g$. The most relevant quantity to be computed
in order to derive term by term at all orders the contributions to
this expansion is the extended generating functional ${\cal
  Z}_{sm}[J,\mu]$ analogous to that of Eq.~(\ref{zfunctextchain}).
After a long but straightforward calculation, one obtains apart from a
trivial overall factor the following expression of ${\cal
  Z}_{sm}[J,\mu]$:
\begin{equation}
{\cal
  Z}_{sm}[J,\mu]=e^{\imath\int_Md^2x\boldsymbol J\cdot\boldsymbol\mu}
e^{-\int_Md^2xd^2y{\cal G}(x,y)\boldsymbol\mu(x)\cdot\boldsymbol\mu(y)}\label{smfinal}
\end{equation}
where
\begin{equation}
{\cal G}(x,y)=\frac g2\Delta_x\delta^2(x,y)
\end{equation}
and $\Delta_x=\frac{\partial^2}{\partial
  x_1^2}+\frac{\partial^2}{\partial x_2^2} $.
\section{Concluding remarks}
In this work a recipe has been presented for constructing the
generating functional of 
constrained stochastic systems with an arbitrary number of degrees of
freedom and general constraints.
The main idea is that the variables to be constrained are allowed to
fluctuate thanks to the addition of the auxiliary noises $\tilde
\nu_\alpha$ and of the fictitious variables $\xi_\alpha$,
$\alpha=1,\ldots,M$. In this way, the conditions
$C_\alpha(\{\boldsymbol R\})$ in Eq.~(\ref{Langevinconstraints}) are
not fixed exactly, but are subjected to stochastic fluctuations. The
original constraints 
(\ref{constraints}) are recovered in the ``rigid'' limit $\eta_\alpha,\tilde
D_\alpha\to 0$. The set of equations (\ref{Langevinequationsbis}) and
(\ref{Langevinconstraints}) which describes the dynamics of both
physical and fictitious degrees of freedom is formally a system of
overdamped Langevin equations. Starting from a system of that kind it is
possible to derive a path integral expression of the related
generating functional by using the standard techniques valid
for stochastic systems without constraints. In the rigid limit
$\eta_\alpha,\tilde D_\alpha=0$, the generating functional
$Z[\boldsymbol J]$ is given in Eq.~(\ref{genfuncfinal}). Remarkably,
with this procedure the jacobian determinant appearing after the
change of variables $\boldsymbol\nu_i,\tilde\nu_\alpha\longrightarrow
\boldsymbol R_i,\xi_\alpha$ remains relatively simple.
For comparison, the standard treatment of constrained path integrals,
see for instance \cite{itzyksonzuber}, produces complicated
jacobian determinants of block matrices. The path integral expressions
obtained in this way are very useful at a theoretical level, because
they allow 
 for example to check if the Parisi-Wu
quantization scheme delivers in the equilibrium limit the desired
generating functional of the theory to be quantized
\cite{Mochizuki,okano}, but are 
very cumbersome in concrete calculations of physical observables.
The method
discussed in this work produces instead relatively simple generating
functionals. In the case of the inextensible chain
in the absence of external interactions, we
obtain for instance the GNL$\sigma$M of Ref.~\cite{FePaVi}, 
which allows the computation of several measurable quantities.
For example,
the dynamical form factor of the chain has been evaluated in
\cite{FePaVi2}. Always in \cite{FePaVi2} it has been estimated how the
fluctuations of the distance between two arbitrary points on the chain
are influenced by physical parameters like the length of the chain and
the relaxation time $\tau$.
The novelty of the present approach with respect to \cite{FePaVi} is the
possibility to add
also the external forces $\boldsymbol f_i$.

Of course, the  expressions of the generating functionals obtained
here are still
complicated due to the presence of the Dirac delta functions that are
needed to impose the constraints.
For this reason, in this work a technique is developed for computing the
 the generating functional of two dimensional
field theories in the presence of a  Dirac delta functions
like that appearing in the GNL$\sigma$M
of Eq.~(\ref{genfuncgnlsm}). 
Thanks to the introduction of the auxiliary field $\boldsymbol
r(t,s)$, the difficulties related to the 
delta function are confined to the computation of the path integral ${\cal
  Z}_2[\boldsymbol k]$ of Eq.~(\ref{z2k}).
 Apart from the boundaries at $t=0$ and
$t=t_f$, Eq.~(\ref{discretizedexact2}) shows that ${\cal
  Z}_2[\boldsymbol k]$ formally consists of the infinite product
over the time index $m$ of terms that are nothing but
the Fourier transform of the probability function of a freely jointed
chain in statistical mechanics. The functional
${\cal
  Z}_2[\boldsymbol k]$ has been evaluated here in the form of a series
converging for small values of the parameter $c$ given of
Eq.~(\ref{cdef}). Despite the fact that Eq.~(\ref{pertexp}) looks like
a perturbative expansion, all terms entering in the expansion
can be computed in closed form. Indeed, 
the main ingredient in Eq.~(\ref{pertexp})
is the
extended generating functional ${\cal Z}[\boldsymbol J,\boldsymbol
  \mu]$, which generates both correlation functions of the physical
fields $\boldsymbol R(t,s)$ and of the auxiliary fields $\boldsymbol
k(t,s)$.
The exact expression of  ${\cal Z}[\boldsymbol J,\boldsymbol
  \mu]$ is given by
Eq.(\ref{extendedgenfun}).

Finally, the techniques presented in this work can be extended to other
systems with constraints. For instance, 
we have seen there are many similarities
between the  two
dimensional 
 nonlinear sigma model and the GNL$\sigma$M.
Despite the initial difference of the constraints,
after the introduction of auxiliary fields
the generating functional of both theories may be computed in the same
way.
In particular, the extended generating functional
${\cal
  Z}_{sm}[J,\mu]$ of the correlation functions of 
the physical fields $\boldsymbol \phi(x)$ and the auxiliary fields
$\boldsymbol k(x)$ for
the nonlinear sigma
model has been exactly derived in Eq.~(\ref{smfinal}).
We hope that in the future it will be possible
to apply these methods also to
the statistical mechanics of polymer interactions in the presence of
topological entanglement \cite{vilgis,vologodskii, marko}. 
In statistical mechanics, in fact, the probability
function of a system of long and flexible polymer chains obeys 
 pseudo-Schr\"odinger equations that, for certain aspects, are
similar to those of Fokker-Planck. A nice physical application 
 could be the
study of the influence of topological constraints on the  interactions in
colloid-polymer  mixtures \cite{chervanyov,ChFe}.
\section{Acknowledgments}

Support from the National Science Foundation of the United States
under Grant No. NSF PHY05-51164 and by the Polish National Center of Science,
scientific project No. N N202 326240 is gratefully acknowledged.

\end{document}